%% file: Vinokurov_NGC7793_P13_donor_WIND.tex
\documentclass[
 aps, pra,
 amsmath,amssymb,
 11pt,
 final,
tightenlines,
 twoside,
 twocolumn,
 nofloats,
nofootinbib,
 superscriptaddress,
showkeys,
showkeywords,
 ]
{revtex4-2}

\usepackage[T2A]{fontenc}
\usepackage[utf8x]{inputenc}
\usepackage[russian,english]{babel}
\usepackage{graphicx}
\usepackage{dcolumn}
\usepackage{bm}

\usepackage[dvipsnames]{xcolor}
\definecolor{myred}{rgb}{0.90, 0.1, 0.1}

\input{maik.rty}

\setcitestyle{authoryear,round,aysep={,},citesep={;}}
\input{sao_cmd_author.tex}

\begin{document}

\selectlanguage{english}



\title{The Nature of the Emission Spectrum of NGC\,7793~P13: modeling the atmosphere of the donor star}

 \author{\firstname{A. S.}~\surname{Vinokurov}}
 \email{vinokurov@sao.ru}
 \affiliation{Special Astrophysical Observatory of the Russian Academy of Sciences, Nizhnij Arkhyz, 369167 Russia} 

 \author{\firstname{A. E.}~\surname{Kostenkov}}
 \email{kostenkov@sao.ru}
 \affiliation{Special Astrophysical Observatory of the Russian Academy of Sciences, Nizhnij Arkhyz, 369167 Russia}
 
\author{\firstname{K. E.}~\surname{Atapin}}
 \affiliation{Sternberg Astronomical Institute, Lomonosov Moscow State University, Moscow, 119992 Russia}

 \author{\firstname{Yu. N.}~\surname{Solovyeva}}
 \affiliation{Special Astrophysical Observatory of the Russian Academy of Sciences, Nizhnij Arkhyz, 369167 Russia}

\begin{abstract}
We continue to study the ultra-luminous X-ray source NGC\,7793~P13 in the optical range. In this work, we are testing the model of a spherically symmetric wind atmosphere of the donor star, previously identified as a B9\,Ia supergiant. The model spectrum has shown good agreement with the observed one at a relatively high mass loss rate of $\dot{M} \approx 6\times10^{-6}\,M_{\odot}\,\text{yr}^{-1}$; other parameters turned out to be close to those expected for late B-supergiants. The increased mass loss rate can be explained by the high rotation velocity of the star. In addition, we have qualitatively demonstrated the effect of X-ray irradiation on the observed spectrum and discuss the fundamental possibility of wind acceleration under conditions of powerful irradiation.
\end{abstract}

\maketitle

\section{INTRODUCTION}

Ultraluminous X-ray sources (ULXs) are point objects that are located outside the centers of galaxies and have X-ray lumiosities greater than $10^{39}$ erg\,s$^{-1}$, which is the Eddington limit for a black hole with a mass of about $10\,M_\odot$. According to current thinking, most ULXs are binaries whose energy release is facilitated by super-Eddington (supercritical) accretion onto neutron stars or stellar-mass black holes \citep[see the reviews by][]{Fabrika2021,King2023,Pinto2023}. Both the theoretical works \citep[e.g.,][]{Shakura1973,Lipunova1999,Poutanen2007,Chashkina2019} and the multiple publications devoted to radiation (magneto)hydrodynamic simulations of the supercritical regime of accretion \citep{Ohsuga2005,Ohsuga2009,Ohsuga2011,Kawashima2012,Takahashi2015,Takahashi2017,Sadowski2016,KawashimaOhsuga2020,Yoshioka2022} predict that the radiation pressure in the central regions of the supercritical accretion disk has to increase the disk thickness and produce powerful high-velocity outflow of matter from its surface (the so-called supercritical disk wind). The recent discovery of emission and absorption features in the X-ray spectra of several ULXs blueshifted by $V_w \approx 0.1\,c$  has confirmed the existence of such outflows  \citep{Pinto2016,Pinto2017,Kosec2018,Pinto2021}. 

As for the optical spectra, having analyzed the spectra of a series of ULXs, \citet{Fabrika2015} concluded that the observed emission lines are most probably formed in the supercritical disk winds, whose physical characteristics should be more similar to those of stellar winds rather than the outflows from the surfaces of geometrically thin accretion disks observed in Galactic X-ray binaries during outbursts. This similarity led us to the idea that the methods developed for modeling of extended atmospheres of stars can also be applied to quantitative analysis of ULX spectra; we attempted to realize this idea in the papers \citet{Kostenkov2020a,Kostenkov2020b,Kostenkov2023}. In the last paper we used the non-LTE code \texttt{CMFGEN} to measure the wind parameters for the ultraluminous X-ray pulsar NGC\,7793~P13~--- a system comprising a neutron star and a B9\,Ia supergiant \citep{Motch2014}. The source has a pronounced emission spectrum, indicating the presence of a powerful wind, much stronger than should be expected from a typical supergiant of this class \citep[not exceeding $5\times10^{-7}\,M_{\odot}\,\text{yr}^{-1}$][]{Crowther2006, Markova2008}, and we therefore deemed it reasonable to consider a two-component model where the main portion of the continuum emission and the absorption lines are formed in the donor while the emission spectrum is formed in the supercritical disk wind. However, due to the small size of the wind photosphere in comparison with the size of the donor, the contribution of the disk wind to the total optical luminosity of the object amounted to less than 10\%. 

This model was able to reproduce the emission line intensities rather well, however, at the same time, a whole set of shortcomings were revealed. The most serious problem arose when reproducing the observed P\,Cyg profiles: we had no options but to ``construct’’ them from two separate components, shifting the absorptions of the donor with respect to the disk wind emissions toward the blue side by a maximally possible for this orbital phase Doppler shift of 80~km\,s$^{-1}$. But even with this shift the depths of the absorption components of some line profiles were underestimated by a factor of 2--3. We have shown that in order to describe the profiles of the H$\alpha$, H$\beta$ and H$\gamma$ lines, as well as the positions of purely absorption lines of the Balmer series, a velocity of 100--250~km\,s$^{-1}$ is required, whereas the metal lines need a lower velocity, down to 20~km\,s$^{-1}$ for Mg\,II~$\lambda4481$. Thus, we have not been able to adequately reproduce within the framework of the combined model not only the P\,Cyg profiles, but also the positions of many absorption lines. Another suspicious result was a similarity of the temperatures of the supercritical disk wind photosphere (the mass loss rate in which was estimated at $\dot M\approx 1.4\times10^{-5}~M_\odot$\,year$^{-1}$) and the donor star: the difference between these temperatures turned out to be no more than 20\%.

These facts altogether suggest that the source of both the emissions and the absorptions is, in fact, the donor star, which, for some reason, produces the wind with the required high $\dot M$ by itself. This idea of the presence of a powerful wind in the donor agrees with the estimate by \citet{El_Mellah2019}, who wondered what mass loss rate the P13 donor should have in order to provide the observed X-ray luminosity if the ``wind Roche lobe overflow'' accretion takes place (they obtained $\dot{M}\simeq10^{-5}~M_\odot$\,year$^{-1}$). One of the reasons that the authors considered this mechanism instead of the usual Roche-lobe overflow in the case of NGC\,7793~P13 is the known problem of instability of the Roche-lobe accretion in systems with high donor-to-accretor mass ratios.

In this work we continue investigating NGC\,7793~P13 and test the possibility of reproducing its optical spectrum in the framework of the donor star atmosphere model with the wind parameters left free. The text of the paper is organized as follows: Section~2 outlines the spectral data used in this work; Sections~3 and~4 describe the method of determining model parameters and the simulation results; Section~5 discusses issues concerning the He\,II~$\lambda\,4686$ emission line which demanded a special treatment; the summary of our results and the discussion of the obtained model parameters are in Section~6.

\section{Observational data}

We used the same NGC\,7793~P13 spectra as in our previous paper \citep{Kostenkov2023}. These observational data were obtained as a part of the NGC\,7793~P13 spectral monitoring program being carried out in 2009--2011 with the VLT telescope using the FORS-2 spectrograph \citep{Motch2011,Motch2014}. For the modelling we used the spectra taken consecutively on November 17, 2009 in the blue (3700--5200\,\AA, GRIS\_1200B+97 grism) and the red (5750--7310\,\AA, GRIS\_1200R+93 grism) wavelength ranges. The spectral resolution is 2.3\,\AA. The object's brightness at the moment of observations \mbox{$V = 20\,.\!\!^{\rm m}0 \pm 0\,.\!\!^{\rm m}01$} was close to the maximum, which occurs when the hemisphere of the star irradiated by X-rays becomes best visible to the observer \citep{Motch2014}. The reasons why we chose these certain observations, as well as the detailed description of the data reduction process are described in the first part of our study devoted to NGC\,7793~P13 \citep{Kostenkov2023}.

\section{METHODS}

Here we consider a model that assumes the formation of the entire optical spectrum of NGC\,7793~P13 (both the continuous component and the spectral lines) in the extended atmosphere of the donor star, the optical emission of the accretion disk is considered negligible. The source of ionization for the wind was the star photosphere whose average effective temperature was enhanced by X-ray heating; the question of the direct irradiation of the wind matter by high-energy quanta is discussed in Sections~5 and~6. The simulations were performed using the {\tt CMFGEN} non-LTE code \citep{Hillier1998}, which solves the radiative transfer equation in the comoving reference frame for expanding atmospheres in spherically-symmetric geometry. A description of the main parameters used by {\tt CMFGEN} is presented in our previous paper, \cite{Kostenkov2023}.

To simplify the model and reduce the amount of needed computation time, for the subsonic region of the wind, we employed the effective scale height of an isothermal  photosphere $h_\text{eff}$ instead of the quasi-hydrostatic wind approximation \citep{Kostenkov2023}. This simplified approach to computing the structure of the lower layers of the atmosphere not yet accelerated enough to become a full-fledged wind was successfully applied to simulating both the dense extended atmospheres of LBV stars \citep{Najarro1997} and the optically thin winds of O stars \citep{Hillier2003}. For the region $v>v_{\rm sonic}$, the dependence of the wind velocity on distance remained within the simple $\beta$-law. Thus the total expression for the velocity was:

\begin{equation}\label{betalaw}
    v(r)=\cfrac{v_0 + (v_{\infty} - v_0)(1 - R_* / r)^\beta}{1 + (v_0/v_{\rm core})\exp[{(R_* - r)/h_{\rm eff}}]}\,,
\end{equation}
where $\beta$ is the exponent parameterizing the wind acceleration with distance from the center of the star; $v_{\rm core}$ is the velocity at hydrostatic radius $R_*$ (in {\tt CMFGEN}, the hydrostatic radius is defined as the inner radius of a model with $\tau_{\rm Ross} \gtrsim 20$), $v_0$ is the parameter determining the gas velocity distribution in the transition zone between the lower layers of the atmosphere and the wind, $v_\infty$ is the terminal wind velocity. The $v_{\rm core}$ value was selected such that the optical depth at the hydrostatic radius remained within \mbox{$20 \leqslant \tau_{\rm Ross} \leqslant 100$}. The quantity $v_{0}$ was adopted equal to $v_{\rm sonic}\approx10$\,km\,s$^{-1}$.

The effective scale height $h_{\rm eff}$ is related to temperature and gravity acceleration at the photospheric distance as:

\begin{equation}\label{heff}
    h_{\rm eff}=1.2 \times 10^{-3}\:\frac{(1 + \gamma)T_{\rm ph}}{\mu (1-\Gamma)g}\,R_\odot,
\end{equation}
where $\mu$ is the average atomic mass in atomic mass units, $\gamma$ is the average number of electrons per atom, $\Gamma$ is the ratio of the total radiation pressure to gravity acceleration $g$. 

The model fitting process was performed primarily by varying the mass loss rate $\dot{M}$ and the photospheric temperature $T_{\rm ph}$ (by $\tau_{\rm Ross} = 2/3$). As for multiple additional parameters, measuring their values directly from the observational data is complicated because of their complex interconnected influence of on the model spectra. For example, the strong dependence of helium and metal line intensities on the microturbulent velocity $v_{\rm turb}$ does not allow one to accurately estimate either the abundances of the corresponding elements, or the temperature of the lower atmospheric layers due to the influence of line blanketing effects on the ionization state of matter \citep{Markova2008_2}. The volume filling factor $f$ parameterizing micro-inhomogeneities in the matter and the exponent $\beta$ in the wind velocity law cannot be unambiguously determined by using only optical observational data \citep{Mokiem2007, Hawcroft2021, Kostenkov2020c}. In some cases it is possible to estimate the terminal wind velocity $v_\infty$ by the half-widths of the forbidden lines formed in outer parts of the wind, for example, such as [N\,II]~$\lambda\,$5755 and some [Fe\,II] lines \citep{Stahl1991, Stahl2001, Gvaramadze2010}, however, such terminal velocity indicators are absent in the optical spectrum of NGC\,7793~P13. Thus, model parameters that are hard to measure were either postulated or varied within the standard limits (see below).

The micro-turbulent velocity $v_{\rm turb}=10$~km\,s$^{-1}$ was chosen to correspond to the lower boundary of the $10$--$20$~km\,s$^{-1}$ interval, obtained from observations of B-type supergiants \citep{Trundle2004, Crowther2006, Markova2008}, similar to the adopted value in the previoгs part of our work. 

The volume filling factor in winds of B supergiants has a high uncertainty. Hydrodynamic wind simulations consistent with the evolutionary models give $f\approx0.05$ \citep{Krticka2024}, which is close to the values obtained in the studies of lines in ultraviolet spectra of O stars, $f\approx0.02$--$0.1$ \citep{Puls2008, Krticka2017, Bouret2021}. At the same time, detailed modeling of the winds of four B supergiants carried out by \citet{Bernini-Peron2023} has shown that they have a more uniform wind with $f \gtrsim 0.5$. We adopted the mean value of the volume filling factor to be $f=0.1$ considering that the wind inhomogeneities originate in the deep layers of the atmosphere $v_{\rm cl}=30~$km\,s$^{-1}$, similar to the winds in O stars \citep{Hillier2003, Bouret2003, Bouret2005}, and the filling factor decreases with increasing distance (starting from $f=1.0$, homogeneous wind) as $f(v)=f_\infty + (1 - f_\infty)\exp[-v/v_{\rm cl}]$ \citep{Hillier1999} with $f_\infty=0.1$. Note that the micro-inhomogeneity of matter in inner and outer parts of the wind may differ significantly depending on the properties of the outflowing material. More accurate estimations should involve combined studies in the optical, infrared, and radio spectral ranges \citep{Puls2006}.

Simulations of optical spectra of B supergiants with non-LTE codes have shown that the values of $\beta$ needed for optimal H$\alpha$ line description fall in the range of 1--3  \citep{Kudritzki1999, Trundle2004, Crowther2006, Markova2008}. However, hydrodynamic computations of dense winds in massive stars with temperatures below the bistable limit predict a rapid increase in the wind velocity $\beta \approx 0.7$--$1.0$ \citep{Vink2018}. As shown by \citet{Petrov2014}, the $\beta$ values obtained as a result of fitting of optical spectra may be overestimated due to the presence of optically thick inhomogeneities in the wind. Thus, based on the results of comparison of hydrogen line profiles in the observed and model spectra for different values of $\beta$ (see details in Section 4), the optimal value was selected to be $\beta=1.0$.

Taking into account the empirical relation $v_{\infty}/v_{\rm esc}\approx 0.9$--$1.4$ for the stars with temperatures of $10$--$17$~kK \citep{Lamers1995, Kudritzki1999, Crowther2006, Markova2008} and the probable donor star mass range $18$--$23~M_{\odot}$ \citep{Motch2014}, we adopted the terminal velocity $v_{\infty}=350$~km\,s$^{-1}$, which corresponds to  $v_{\infty}/v_{\rm esc}=1.3$ for $M=20\,M_{\odot}$ and $R_*=100\,R_\odot$. This value allowed us to adequately reproduce both the observed velocities of absorption components and the half-widths of the Balmer series emission lines for the fixed $\beta=1.0$. 

The chemical composition of the donor star model atmosphere (${\rm He/H}=0.20$ by the number of atoms, \mbox{$X_{\rm C}/X_{\odot}\approx0.1$} in mass fraction, $X_{N}/X_{\odot}\approx1.0$, $X_{O}/X_{\odot}$=0.15) was identical to the chemical composition of B supergiants \citep{Crowther2006} with account for the reduced metallicity $Z=0.5\,Z_\odot$ of the NGC\,7793 galaxy \citep{Pilyugin2014}. The abundances of heavier-than-oxygen elements included in the model (Ne, Mg, Si, Ca, Fe) were adopted at $Z=0.5\,Z_\odot$. 

We selected $T_{\rm ph}=13$~kK as an initial estimate for the photospheric temperature, this value was obtained in our previous simulations within the combined model \citep{Kostenkov2023}, it takes into account the heating of the star's surface by X-ray radiation. The $\dot{M}$ value for the first stages of the modeling was selected to be in accordance with the equivalent widths of the Balmer series emissions (mainly H$\alpha$) and the He\,I triplets \citep{Kostenkov2020a}. Deviations of the modeled intensities of these lines from the observed ones were considered acceptable if they did not exceed 15\%. Variations of the photospheric temperature within $\Delta T_{\rm ph}=\pm3$~kK do not noticeably affect the $\dot{M}$ estimates because the wind ionization state does not change much in between the bistable transitions taking place at temperatures of about 20~kK and 10~kK, respectively \citep{Groh2011, Petrov2016}.  

Further, in order to adjust the photospheric temperature $T_{\rm ph}$, we calculated a model grid in the temperature range from 10~kK to 15~kK with a step of 500~K. As in our paper \citet{Kostenkov2023}, the choice of the optimal $T_{\rm ph}$ was based on the joint fit of the He\,I and Mg\,II~$\lambda\,$4481 lines, as well as on the depths of the absorption components of the Balmer series lines and the He\,I~$\lambda\,$6678/$\lambda\,7065$ singlet-to-triplet emission intensity ratio. The temperature $T_{\rm ph}$ was varied by changing the star luminosity which allowed us to make hydrostatic radii about the same in the calculated models and, as a consequence, identical wind density distributions for the same mass loss rates.

After deriving the optimal mass loss rate $\dot{M}$ and effective temperature $T_{\rm ph}$, the effective scale height of the atmosphere $h_{\rm eff}$ was adjusted using the depths of the absorption components of H$\delta$, H$\epsilon$ and higher order Balmer series lines, where the contribution of the wind-formed emissions is minimal. The of $h_{\rm eff}$ on the model spectrum is similar to $g$ in the case of a quasi-hydrostatic approximation for the lower atmospheric layers. The simulated spectra were smoothed using the \texttt{rotBroad} procedure of the \texttt{PyAstronomy} package to imitate the rotation of the star with a projected velocity of 50~km\,s$^{-1}$.

At the final stage, the luminosity and radius of the model were corrected with account for the apparent magnitude of the object \mbox{$V=20\,.\!\!^{\rm m}0$}, distance to the NGC\,7793 galaxy 3.7\,Mpc \citep{RadburnSmith2011} and interstellar extinction \mbox{$A_V = 0\,.\!\!^{\rm m}2$} \citep{Motch2014} using the reddening curves from \citet{Fitzpatrick1999}. The final model parameters determining the wind density were then scaled in accordance with the $L\propto R^{2}$ and $\dot{M}\propto R^{1.5}$ relations \citep{Najarro1997}. In order to reach better agreement between the observed and modeled spectra, the mass loss rate was corrected after changing the model radius.

\section{RESULTS}

\begin{table*}
\setlength{\tabcolsep}{4pt}
\caption {Basic model parameters: $L$~--- luminosity of the donor star on the assumption that its whole surface is heated to $T_{\rm ph}$; $\dot{M}$~--- wind mass loss rate; $R_{\rm ph}$~--- photospheric radius; $T_{\rm ph}$~--- photospheric temperature; $R_*$~-- hydrostatic radius corresponding to the optical depth $\tau_{\rm Ross}\approx75$ in our model; $T_{*}$~--- temperature at $R_*$; $h_{\rm eff}$~--- effective scale height of the atmosphere; $v_{\infty}$~--- terminal wind velocity; $\beta$~--- exponent of the wind velocity law; $f$~--- volume filling factor; $v_{\rm cl}$~--- velocity from which the wind ceases being homogeneous; $v_{\rm turb}$~--- microturbulent velocity of the star's atmosphere.}
\label{tab_par}
\medskip
\begin{tabular}{l|c|c|c|c|c|c|c|c|c|c|c|c}
\hline
\multirow{2}{*}{Parameters}&$L$, &$\dot{M}$, &$R_{\rm ph}$, &$T_{\rm ph}$,&$R_*$, &$T_{*}$, &$h_{\rm eff}$,  &$v_{\infty}$, &\multirow{2}{*}{$\beta$} &\multirow{2}{*}{$f$}&$v_{\rm cl}$, &$v_{\rm turb}$,  \\
& $L_\odot$ & $M_{\odot}\,\text{yr}^{-1}$ & $R_\odot$ & kK& $R_\odot$& kK& $R_*$ & km\,s$^{-1}$ & & & km\,s$^{-1}$ & km\,s$^{-1}$ \\\hline
Values& $3.3\times10^5$&$5.8\times10^{-6}$&$105$&$13.5$ &$101$ &$13.7$& $0.01$& $350$& $1.0$ & $0.1$& $30$& $10$ \\
\hline
\end{tabular}
\vspace{0.2cm}
\end{table*}

\begin{figure*}[t]
     \centering
         \includegraphics[width=\linewidth]{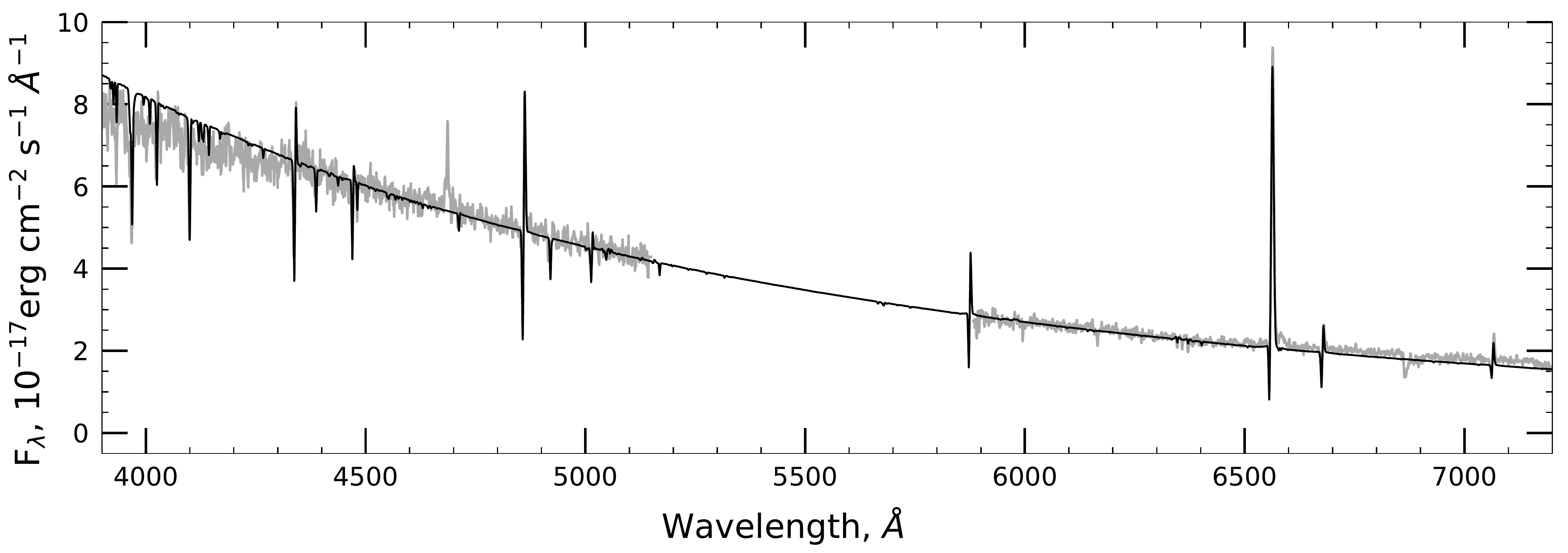}
        \caption{Observed spectrum of NGC\,7793~P13 (grey) and best-fitting model of the donor star's wind spectrum (black).}
        \label{fig:ngc7793_model_flux}
\end{figure*}

\begin{figure*}[t]
     \centering
         \includegraphics[width=\linewidth]{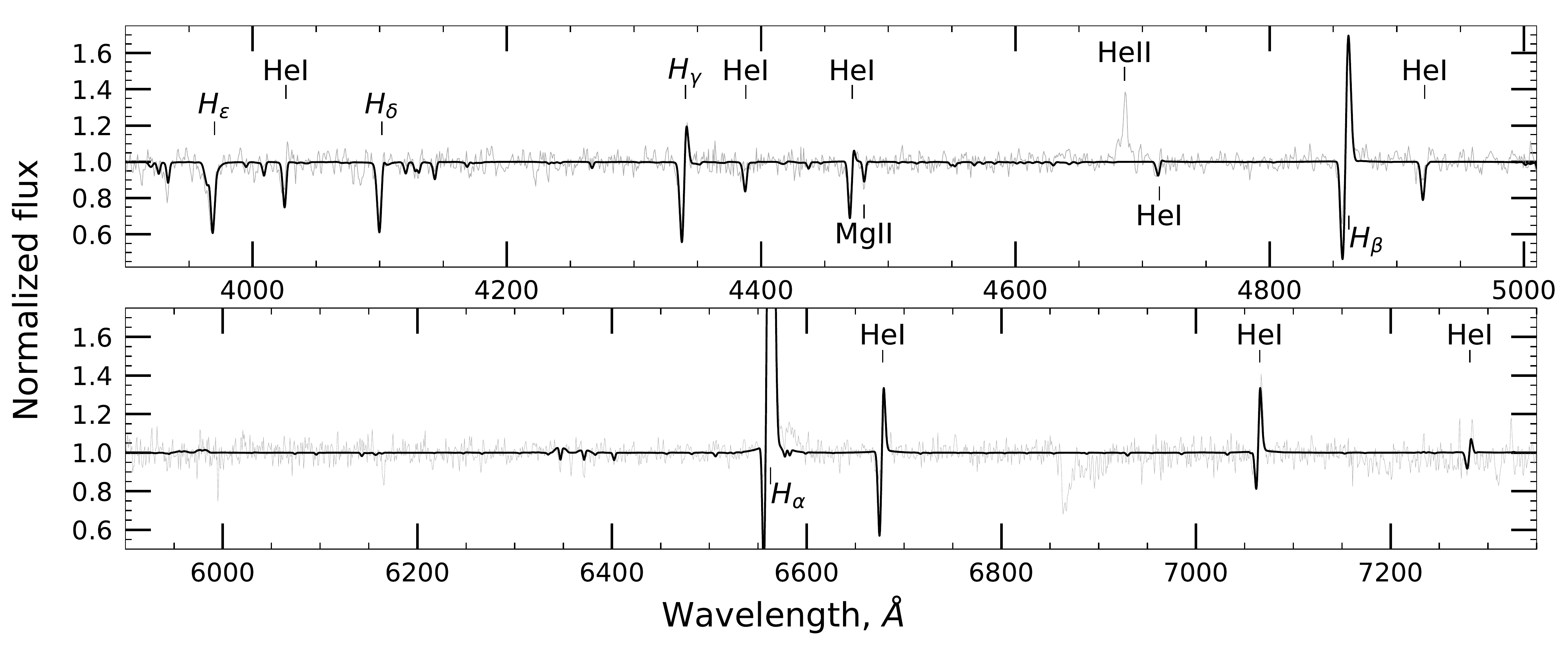}
        \caption{Normalized form of the observed and model spectra of NGC\,7793~P13 shown in Fig~\ref{fig:ngc7793_model_flux}.}
        \label{fig:ngc7793_model_norm}
\end{figure*}

\begin{figure*}[t]
     \centering
         \includegraphics[width=\linewidth]{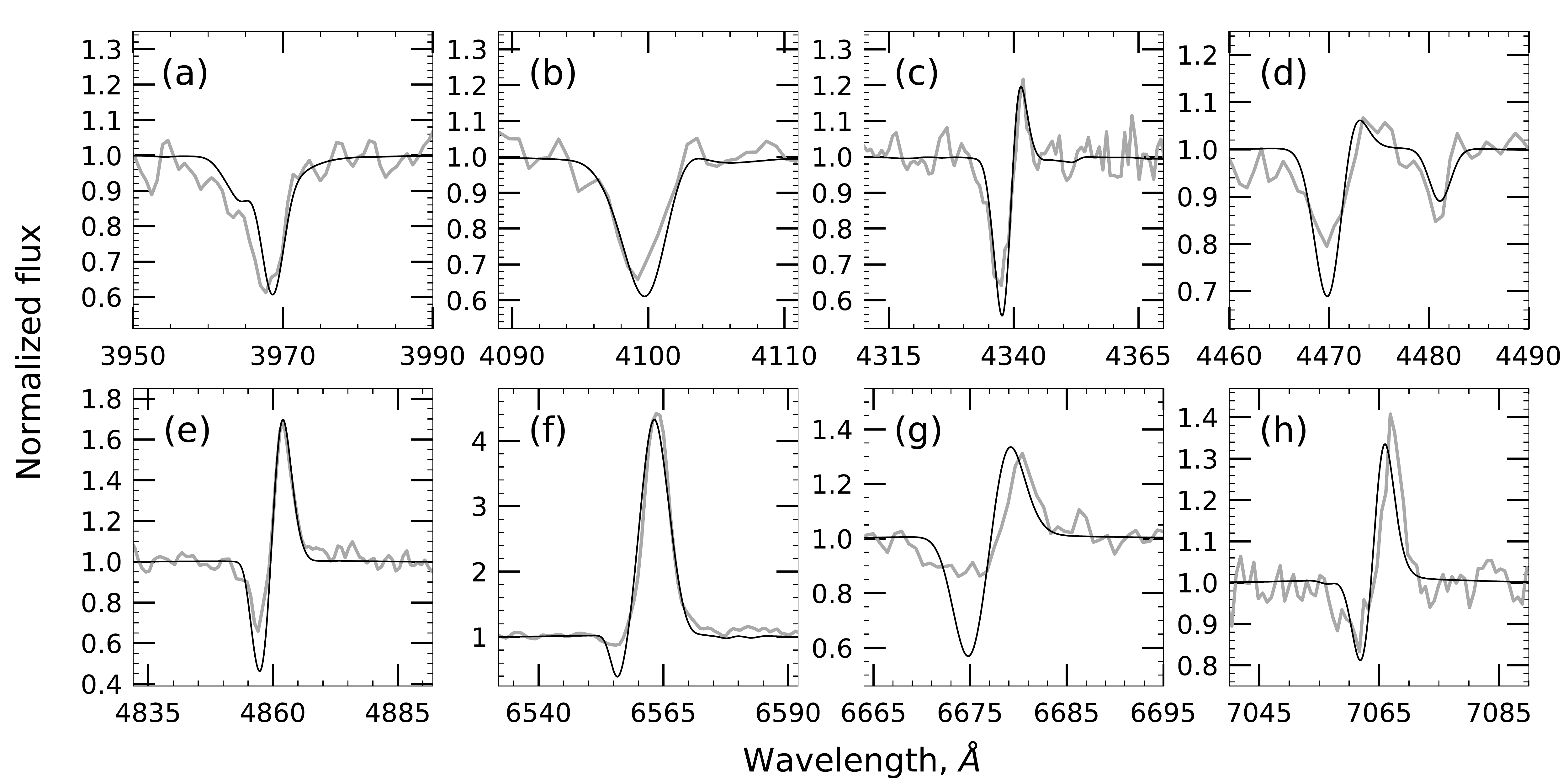}
        \caption{Comparison of the profile shapes for selected spectral lines in the observed (grey) and model (black) spectra: (a) H$\epsilon$; (b) H$\delta$; (c) H$\gamma$; (d) He\,I~$\lambda\,4471$, Mg\,II~$\lambda\,4481$; (e) H$\beta$; (f) H$\alpha$; (g) He\,I~$\lambda\,6678$; (h) He\,I~$\lambda\,7065$.}
        \label{fig:ngc7793_model_lines}
\end{figure*}

\begin{figure}[t]
     \centering
         \includegraphics[width=\linewidth]{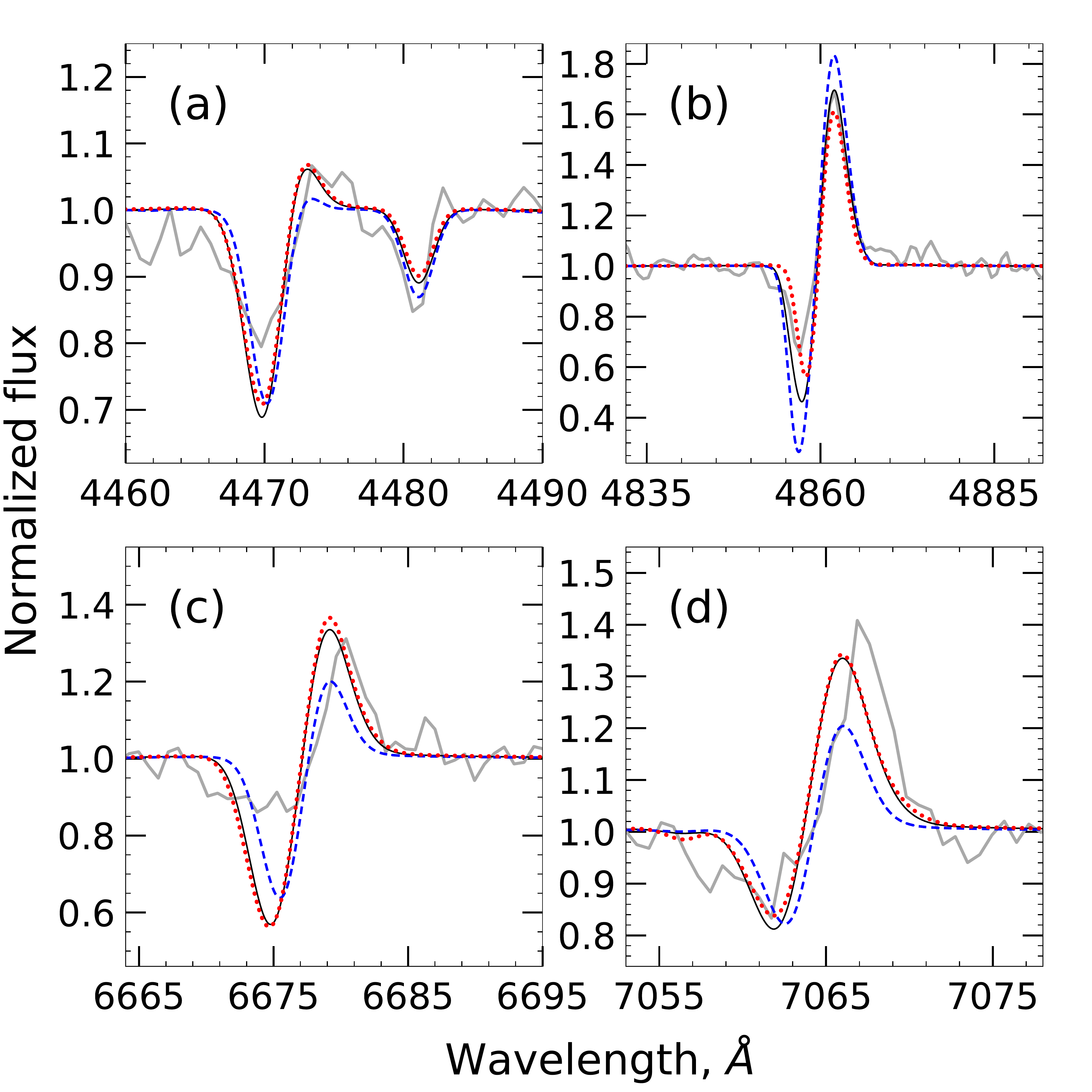}
        \caption{Comparison of the profile shapes of some spectral lines for various photospheric temperatures: $T_{\rm ph}=12.5\,$kK (dashed blue curve), $T_{\rm ph}=13.5\,$kK (solid black curve) and $T_{\rm ph}=14.5\,$kK (dotted red curve). The normalized observed spectrum is shown in grey. The panels show the following spectral lines: He\,I~$\lambda\,4471$, Mg\,II~$\lambda\,4481$ (a), H$\beta$ (b), He\,I~$\lambda\,6678$ (c) and He\,I~$\lambda\,7065$ (d).
}
        \label{fig:ngc7793_model_lines_diff_temps}
\end{figure}

The fitting of the simulated spectra to the observed one was performed `by eye', during this process we tried to reproduce equally well both intensities and half-widths of the emission lines and depths and shifts of the absorption components of various elements. After had calculated over 100 models with different $T_{\rm ph}$, $\dot{M}$, $\beta$, $v_\infty$ values we became able to reach a rather good fit with the exception of some individual lines. Figures~\ref{fig:ngc7793_model_flux} and \ref{fig:ngc7793_model_norm} show the model spectrum in absolute flux units and in normalized form, respectively, and Figure~\ref{fig:ngc7793_model_lines} additionally shows the profiles of the most important lines. Table~1 presents the main model parameters. As in \citet{Kostenkov2023}, the quantities in the table are presented without uncertainties because of the absence of good numerical criteria of their determination, computational difficulties and complex interconnection between the parameters.

During the optimal model selection we studied the effect of the velocity law index $\beta$ (within 1--3) on the final model parameters and spectral line profiles. An increase of $\beta$ to 2--3 expands the formation zone of the Balmer series lines and, at the same time, shifts this region towards lower velocities, which leads to an increase in the line widths for a fixed $v_\infty$ \citep[e.g.,][]{Najarro1997}. Thus, to preserve the line profile at $\beta>2$, one would need to significantly increase the terminal velocity, up to 500--700~km\,s$^{-1}$, which is multiple times higher than $v_\infty$ expected for the P13 donor. Also, at $\beta>2$, the blue shift of the absorption components of hydrogen lines and He\,I becomes insufficient because of slower wind acceleration in the lower layers of the atmosphere. That is why we accepted the standard value $\beta=1.0$, which allows reproducing at high wind mass loss rates (see below) both the velocities of absorption lines for various ions and the widths of the hydrogen emission lines at a terminal velocity value of $v_\infty=350$\,km\,s$^{-1}$ not beyond the empirical relations. 

The effective atmospheric scale height $h_{\rm eff}$ also affects the velocity distribution of the outflowing matter near the surface of the star and, therefore, the depth and shift of the absorption components of the line profiles. The optimal value $h_{\rm eff}=0.01$ was selected by fitting the depths of the hydrogen absorptions in the blue part of the spectrum. For the mid-range of the possible donor mass interval of $18$--$23\,M_\odot$ \citep{Motch2014}, the given effective scale height corresponds to $\log{g}=1.71$, which is somewhat lower than the theoretical value for an isolated B9\,Ia star \citep{Straizys1981}.

Minimization of the discrepancy of the helium and metal absorption depths between the simulated and observed spectra was carried out primarily by changing the temperature; the equivalent widths of hydrogen emissions were used to determine the second fundamental parameter of the wind~--- the mass loss rate. Also, both the mass loss rate and, to a lesser extent, the temperature along with the velocity law parameters affected the profiles and positions of many other lines in the spectrum. Thus the model with $T_{\rm ph}=13.5$~kK and $\dot{M}=5.8\times10^{-6}\,M_{\odot}\,\text{year}^{-1}$ was accepted as the most optimal choice among all the considered options.

Varying $T_{\rm ph}$ has a multidirectional effect on lines in the model spectrum, however, there is no solution that fully fits the observations, which is demonstrated in Figure~\ref{fig:ngc7793_model_lines_diff_temps}. The models shown in the figure are computed for different luminosites and, as a consequence, effective temperatures at fixed mass loss rate \mbox{$\dot{M}=5.8\times10^{-6}\,M_{\odot}\,\text{year}^{-1}$} and radius $R_*=101\,R_\odot$. As is evident from the figure, the model with the photospheric temperature 12.5~kK is the best fit for the Mg\,II~$\lambda\,4481$ absorption line, however, using such a temperature leads to significantly underestimated emission components of the He\,I~$\lambda\,6678$, $\lambda\,7065$ lines. The H$\beta$ and H$\alpha$ absorption depths are overestimated in all the models, however, this effect is most conspicuous at low temperatures. Note that the mass lost rate at \mbox{$T_{\rm ph}=12.5$}~kK and \mbox{$T_{\rm ph}=14.5$}~kK should be adjusted within 15\% in order to conserve the hydrogen emission equivalent widths but in the figure we show pure temperature influence on the model spectra. An increase of the photospheric temperature to \mbox{$T_{\rm ph}=14.5$}~kK makes the Mg\,II~$\lambda\,4481$ line weaker but does not change much the equivalent widths of the He\,I~$\lambda\,6678$, $\lambda\,7065$ lines, which remain almost the same as in the best-fitting model with $T_{\rm ph}=13.5$~kK. A further increase in temperature leads to an inversion of the intensity ratio of the helium He\,I~$\lambda\,6678$ singlet and He\,I~$\lambda\,7065$ triplet due to difference in base level depopulation mechanisms \citep{Siviero2003}, however, that is not what is observed in the NGC\,7793~P13 spectrum.

Also, we must note that none of our computed models describes the emission line of ionized helium He\,II\,$\lambda\,4686$. In the absence of extreme conditions such as high turbulence \citep[e.g., as in][]{Kostenkov2023}, the minimum temperature required to produce the ionized helium He\,II~$\lambda\,4686$ line in a standard stellar wind model is 23--25~kK. The optimal temperature $T_{\rm ph}=13.5$~kK that we obtained~--- similar to the results of \citet{Kostenkov2023}, but about 2500~K higher than in \citet{Motch2014}~--- has already taken into account the X-ray heating of the photosphere that should be well visible at this phase of the orbital motion, so we do not reasons for further increasing the photosphere temperature. However, in addition to heating the star's surface, the X-ray quanta should also excite the wind matter itself. Below we show that the accounting for this phenomenon can lead to the appearance of a strong He\,II~$\lambda\,4686$ emission line even in our relatively cool models.

\section{ACCOUNTING FOR DIRECT IRRADIATION OF WIND MATTER BY THE X-RAY SOURCE}

\begin{figure*}[t]
     \centering
         \includegraphics[width=\linewidth]{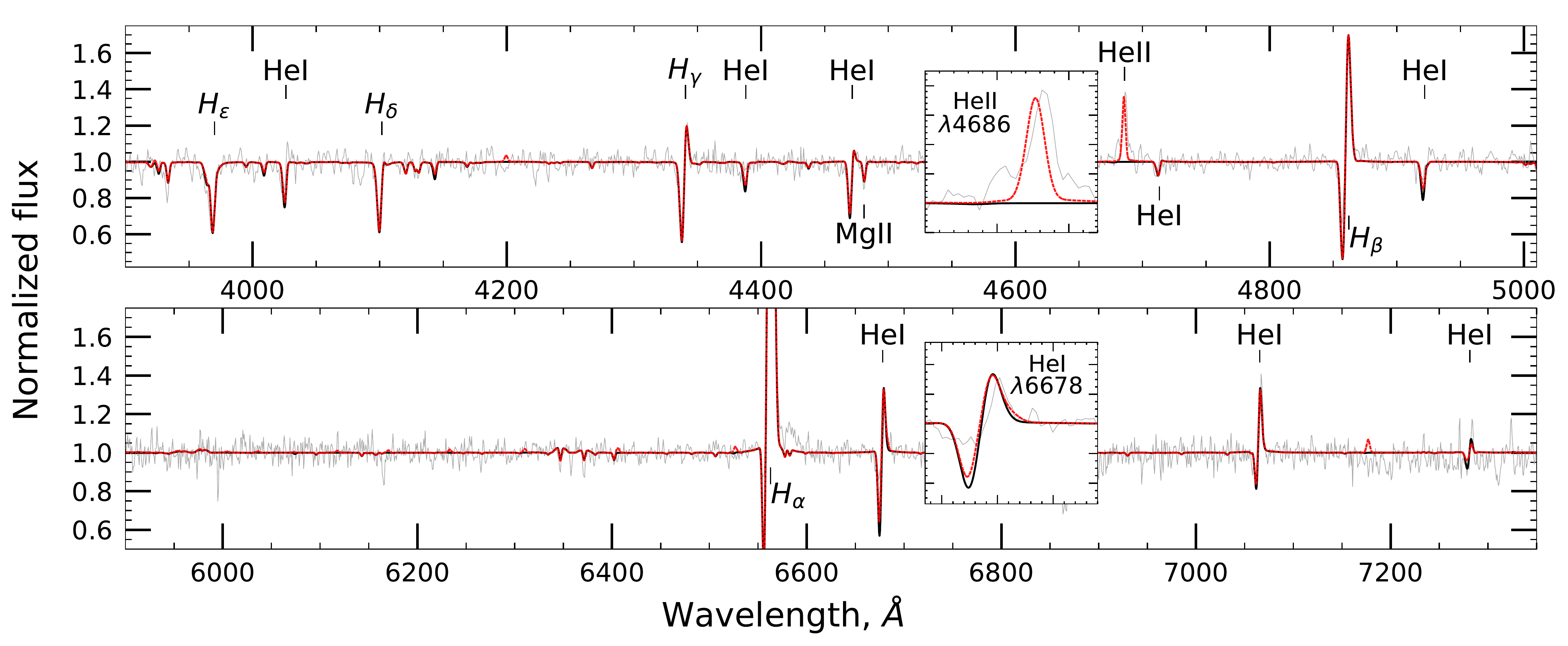}
        \caption{Observed spectrum of NGC\,7793~P13 (grey), best-fitting simulated spectrum of the donor star wind (black, the same as in Figure~\ref{fig:ngc7793_model_norm}) and the same model additionally accounted for irradiation of the wind matter by an X-ray source of $L^X_{\rm total}\approx10^{37}\,$erg\,s$^{-1}$ (red).}
        \label{fig:ngc7793_model_flux_with_xray}
\end{figure*}

A detailed computation of the effects of stellar wind irradiation by X-ray quanta should involve combined radiation-hydrodynamic calculations in 3D geometry. However, due to the high complexity of such computations, at the current stage of development of extended stellar atmosphere modeling techniques, the X-ray irradiation of stellar winds is accounted for mainly by modifying existing spherically symmetric non-LTE models. In particular, \citet{Sander2018} who calculated stellar models with the \texttt{PoWR} code \citep{Grafener2002, Hamann2003},
imitated the X-ray irradiation by adding an extra source of free-free emission corresponding to an optically thin spherical layer of hot plasma somewhere in the wind. Similar methods are used in \texttt{CMFGEN} models for adding the emission of shock waves \citep{Hillier1998}. We have however used an alternative approach, it was presented in the papers \citet{Krticka2012, Krticka2018, Krticka2022} studied wind dynamics in close binary systems. First, the average intensities of the radiation field corresponding to the optimal model described in the previous section were increased with account for the additional contribution of the high-energy quanta. Then, for this modified radiation field, we obtained new level populations as the solution of the statistical equilibrium equations, but the parameters of the optimal model were no longer adjusted. The additive to the average intensity was:
\begin{equation}
    J^X_\nu=\cfrac{L^X_\nu}{16\pi^2d^2}{\rm e}^{-\tau_\nu(r)},
\end{equation}
where $L^X_\nu$ is the monochromatic X-ray luminosity normalized to $L^X_{\rm total} = \int L^X(\nu) {\rm d}\nu$; the X-ray flux was uniformly distributed in the range from the He\,II ionization threshold (about 0.06~keV) to 10~keV, which approximately corresponds to the X-ray spectrum shape of NGC\,7793~P13 based on the results of \citet{Motch2014};
$d$ is the distance from the X-ray source to a certain point in the wind; $\tau_\nu(r)$ is the optical depth at the frequency $\nu$ at that point in respect to the X-ray source defined as:
\begin{equation}
    \tau_\nu(r)= \left\lvert\int\limits_r^D f(r') \chi_\nu(r') {\rm d}r' \right\rvert,
\end{equation}
where $f(r)$ is the volume filling factor, $\chi_\nu(r)$ is the opacity of matter at the frequency $\nu$, $D=9.4\times10^{12}$~cm is the distance between the binary components in periastron for an eccentricity of $e=0.24$ \citep{Furst2021}, the distances are related as $d=|D-r|$. 

The resulting model spectrum that takes into account donor wind irradiation by X-ray quanta is presented in Figure~\ref{fig:ngc7793_model_flux_with_xray}. We found that the luminosity of the source of X-ray irradiation needed to optimally reproduce the most intensive narrow He\,II~$\lambda\,4686$ emission line component is only $L^X_{\rm total}\sim 10^{37}~$erg\,s$^{-1}$, which is 1--3 orders of magnitude lower than the real X-ray luminosity of the object \citep{Motch2014}. Also we obtained the He\,II~$\lambda\,4686$ line shifted toward bluer wavelengths in the model relative to the observed spectrum, but the widths of the narrow line components are in good agreement. The relatively weak broad underlying base may be related to the accretion disk around the neutron star. In addition to producing the ionized helium line, X-ray irradiation of the wind also decreased the depth of the He\,I~$\lambda\,6678$ neutral helium singlet absorption, however, the discrepancy between the observed and model line profiles remains significant. 

An increase in distance between the X-ray source and the star to $D=1.3\times10^{13}$~cm, which corresponds to the radius of the circular orbit that saves the other parameters of the binary unchanged, makes the required X-ray luminosity higher by an order of magnitude, $L^X_{\rm total}\sim 10^{38}~$erg\,s$^{-1}$. As a result, the narrow He\,II~$\lambda\,4686$ component profile becomes broader by approximately a factor of 1.5 in comparison with the previous version of the model (the line flux increases correspondingly), where the source was located in the periastron of an elliptical orbit significantly closer to the star's surface, but the intensity of the He\,I~$\lambda\,6678$ line emission component, as well as the depth of its absorption component decreases by almost a factor of two; additionally, the absorption component shifts toward smaller velocities. Probably, the key to the problem of reproducing these lines lies in rejecting the spherical geometry and taking into account the non-central position of the X-ray source.

\section{DISCUSSION}
In this work we continued investigating the nature of the optical emission spectrum of the ultraluminous X-ray pulsar NGC\,7793~P13. Whereas in our previous paper \citep{Kostenkov2023} we considered a model where the B9\,Ia-type donor star did not produce any emission lines due to the weakness of its own outflows and all the emissions came from the wind of the supercritical accretion disk (the two-component model), now the donor has a high mass loss rate and can produce the observed optical spectrum by itself. The new model has shown a number of advantages in comparison with the two-component one. In particular, despite the smaller number of degrees of freedom, it allowed us to obtain the P\,Cyg profiles of hydrogen and neutral helium lines in a most natural way, as well as to reproduce the depths and positions of the absorption lines, which we were unable to do by adding up spectra of the standard B9\,Ia star and the supercritical disk wind \citep{Kostenkov2023}. Some remaining inconsistencies in the shapes of individual lines may be related to a possible asymmetry of the stellar wind, accounting for which is beyond the scope of this paper. The obtained basic parameters of the donor (Table~\ref{tab_par}): luminosity, radius of the photosphere, effective temperature~--- are close to those expected for a late-type B supergiant and practically do not differ from the values obtained for the two-component model \citep[Table~1 in][]{Kostenkov2023}. Thus, a special discussion is needed only for the obtained value of the mass loss rate in the star's wind and issues related to irradiation of the donor by the X-rays coming from the pulsar.

\subsection{Irradiation of the donor star and collimation of the X-ray emission of the pulsar}

In the two-component model, we achieved the required strength of the He\,II~$\lambda\,4686$ ionized helium line by increasing the microturbulent wind velocity to $v_{\rm turb}=50~$km\,s$^{-1}$. This approach could be considered appropriate for the case of supercritical disk winds \citep{Kobayashi2018}, but such a high turbulent velocity is not observed in the atmospheres of relatively cool stars \citep{Markova2008, Trundle2004, Crowther2006}. Therefore, in the current model, we had to take into consideration the fact that the pulsar's X-ray emission not only heats the photosphere of the star, but also ionizes the wind matter.  

We discovered that, depending on the distance between the X-ray source and the donor, the X-ray luminosity required to reproduce the observed He\,II~$\lambda\,4686$ line intensity should be within the interval of $10^{37} - 10^{38}~$erg\,s$^{-1}$. This may seem surprising because these estimates are 2.5--3 orders of magnitude lower than the (isotropic) X-ray luminosity of the system, both the one known from direct X-ray observations and the one required for explaining the observed variations in the visual band occurring as a consequence of donor heating \citep{Motch2014}. Moreover, the irradiation of the donor wind by an isotropic X-ray source with $L^X_{\rm total}\sim10^{40}$~erg\,s$^{-1}$ must switch the wind to a high ionization state which would make its acceleration impossible, and the wind became suppressed\footnote{New modeling results point to a potential possibility of conserving the donor wind at X-ray luminosities up to $10^{41}$~erg\,s$^{-1}$ in the case of a strongly pronounced inhomogeneous structure of the outflowing matter, but only in systems with a massive accretor \citep{Krticka2018, Krticka2022}. Also, the so-called thermally-driven winds or excited winds may appear in the case of very strong heating \citep{Arons1973,Basko1977}. } \citep{StevensKallman1990}.

We suppose that this inconsistency can be overcome if some parts of the accretion disk obscure the X-ray emission of the central source from the donor star. In particular, many authors \citep[see reviews by][]{Fabrika2021,King2023} discuss the collimation of the X-ray emission of ULXs in geometrically thick accretion disks, where the ``empty'' space above the disk surface in its central parts forms a relatively narrow funnel which releases most of the X-ray quanta \citep{Shakura1973,Lipunova1999,Poutanen2007,Kawashima2012,Sadowski2016,Takahashi2017,Yoshioka2022}. As a result, the main part of the X-ray flux should be emitted within the angle $\theta_f$ from the funnel axis; it should, on the one hand, decrease the estimate of the required X-ray luminosity compared to the isotropic case and, on the other hand, allow only a portion of the donor wind to be illuminated.

Estimates of the apparent luminosity boost produced by such a funnel (characterized by the so-called ``beaming factor'', $B$) depend on the opening angle of the funnel and the details of its geometry. For example, \citet{Dauser2017} having studied the ultraluminous pulsar NGC\,5907~X-1 drew the conclusion that the observed X-ray light curve of this object can be explained if this ULX has a very narrow funnel with an opening angle of $\theta_f \lesssim 10^\circ$ and high $B \sim 100$. {On the other hand}, the simulations by \citet{Mushtukov2021} has shown that, in the general case, high beaming factors contradict the large fraction of pulsating emission typical for ultraluminous X-ray pulsars, but values of $\sim 3-5$ still appears to be acceptable \citep{Takahashi2017, Mushtukov2021}.

\begin{figure}[t]
     \centering
        \includegraphics[width=\linewidth]{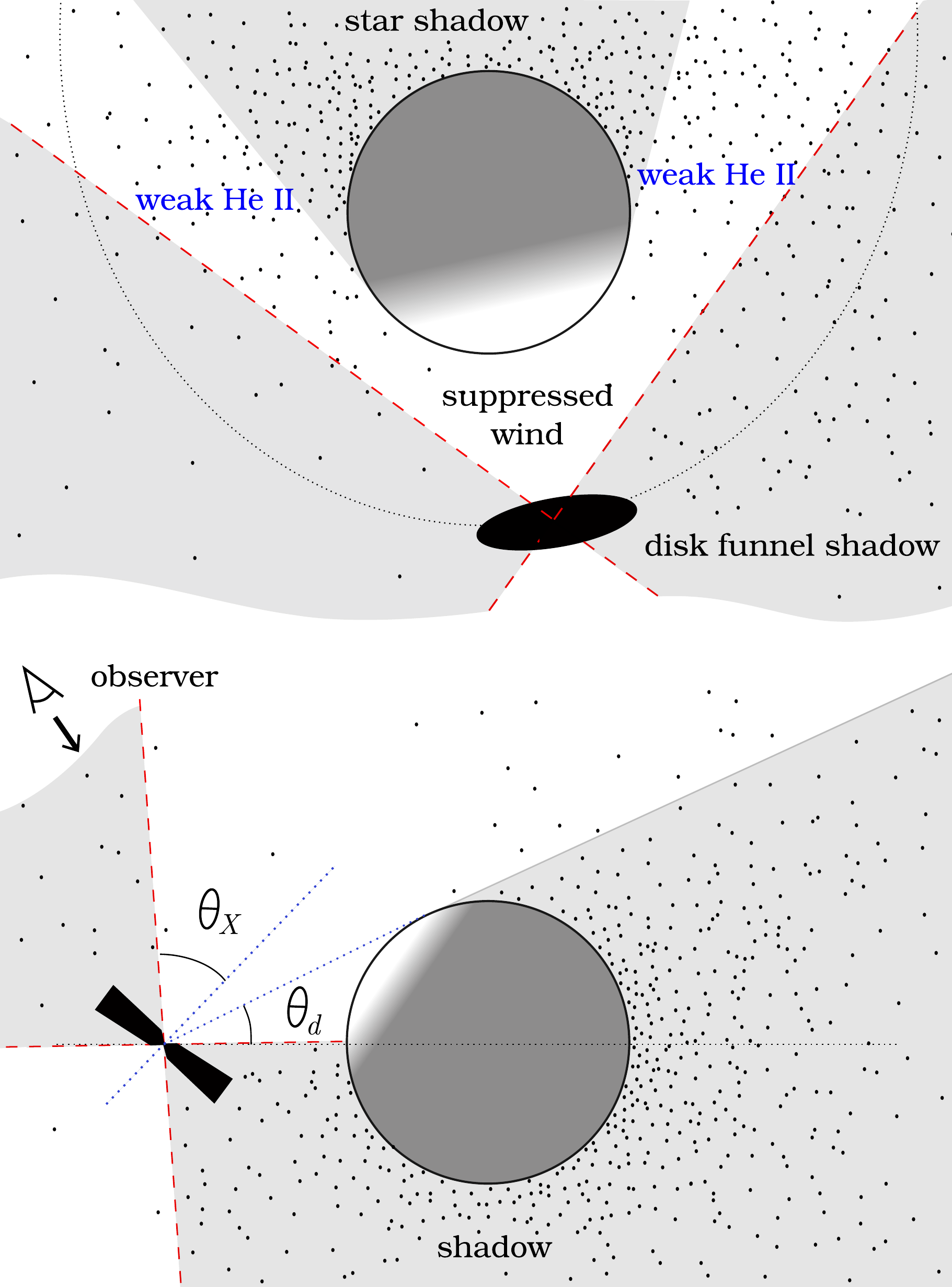}
        \caption{
        A possible configuration that the NGC\,7793~P13 system could have at the moment when the optical spectrum simulated in this work was obtained \citep[the orbital phase is $\Phi=0.04\pm0.03$,][]{Kostenkov2023}. The upper panel shows the view of the system from the pole, the bottom panel from the orbital plane. The X-ray emission of the pulsar is released within a cone with some half-opening angle $\theta_X$. The Earth observer sees the star surface irradiated by X-ray quanta (the star is visible from the neutron star at an angle of $\theta_d$) but may not fall into the X-ray radiation cone if the cone axis is tilted with respect to the normal to the binary plane and $\theta_X$ is very big. Most of the stellar wind is constantly located in the X-ray shadow and maintains a low degree of ionization (see text for details).}
        \label{fig:ngc7793_scheme}
\end{figure}

To summarize the above, we can propose the scenario illustrated in Figure~\ref{fig:ngc7793_scheme}. The funnel axis can be tilted to the orbital plane, resulting in the donor star crossing the X-ray emission cone twice per orbital period. In these moments, the X-ray radiation will effectively heat the photosphere of the star, leading to an increase in its luminosity in the optical and UV spectral ranges, but the wind from the irradiated hemisphere will be suppressed. However, the other parts of the wind: those are outflowing from the far side of the star \citep{Blondin1994} or located outside the radiation cone, will survive; they should remain cool and continue accelerating, what, in principal, could be sufficient to provide the observed spectrum of NGC\,7793~P13. Subsequent orbital motion of the neutron star takes the donor out of the radiation cone, as a results, the wind resumes its outflowing from the star surface, and optical luminosity diminishes due to the cessation of the heating. Figure~\ref{fig:ngc7793_scheme} shows schematically the system configuration at the moment of the spectral observations analyzed in this paper: the donor star is behind the X-ray source, the X-ray radiation cone is directed toward the star.

If the funnel axis is tilted, there is a possibility that the Earth observer does not fall into the radiation cone illuminating the donor star even when the funnel opening angle is relatively wide and the orbit inclination angle is $i\gtrsim20^\circ$ \citep{Motch2014,Furst2018}, see the lower panel in Figure~\ref{fig:ngc7793_scheme}. At these moments, the apparent X-ray luminosity of the system is reduced which agrees well with the observational fact that the optical brightness of the system reaches its maxima precisely during the prolonged minima in the X-ray light curve (see the light curves in \citet{Furst2021}): the apparent X-ray luminosity (recalculated to isotropic emission) goes down to about $5\times10^{37}$~erg\,s$^{-1}$, but the luminosity required to provide the donor heating still remains at a level no lower than $10^{39}$~erg\,s$^{-1}$ \citep{Motch2014}. Besides the orbital variations, the system also exhibits secular changes in both the mean optical and X-ray fluxes and the nature of the variability itself \citep{Furst2021}. This may be a result of either the disk/funnel precession or variations of the opening angle of the funnel $\theta_f$ \citep{Furst2021}. The idea of the collimation of X-ray radiation by a precessing disk has also been mentioned in one of the recent talks by C.~Motch\footnote{\url{https://www.cosmos.esa.int/documents/1518557/1518574/motch.pdf}}, where it was used as an alternative explanation of one of the main features of the P13 optical light curve~--- the narrower maximum in comparison to the minimum, which was previously explained by the ellipticity of the system orbit \citep{Motch2014}. Both the first and the second mechanisms allow decreasing the actual energy release and limiting the direction of X-ray quanta propagation in space, which is precisely what is needed to protect the wind. 

In order to make the donor be able to fall into the radiation cone at the orbital and precession phases illustrated in Figure~\ref{fig:ngc7793_scheme}, the inclination angle of the funnel to the orbital plane should exceed $90^\circ-(\theta_X+\theta_d)$, where $\theta_X\gtrsim\theta_f$ is the half-opening angle of the X-ray radiation cone and $\theta_d$ is equal to half the angle at which the donor is visible from the neutron star. Since $\theta_X$ should be $\approx30^\circ$--$50^\circ$ for the reasonable beaming factors and $\theta_d$ is approximately $34^\circ$ for the donor radius $105\,R_\odot$ derived from the model and a semi-major axis of the orbit of $1.3\times10^{13}$~cm, the angle between the normal to the orbit and the funnel axis should be greater than $10^\circ$--$30^\circ$. These values do not look unrealistic because, for example, the well-known Galactic superaccretor SS\,433 precesses with an angle of about $20.9^\circ$ \citep{Eikenberry2001}.

\subsection{Model parameters: interpretation of the high mass loss rate in the donor star wind}

The mass loss rate in the donor wind obtained in this work, $\dot{M}\approx6\times10^{-6}\,M_{\odot}\,\text{yr}^{-1}$, exceeds by more than an order of magnitude the values typical for B supergiants, \mbox{$(1$--$5)\times10^{-7}\,M_{\odot}\,\text{yr}^{-1}$} \citep{Crowther2006, Markova2008, Markova2008_2, Trundle2004}. The latter rates are insufficient for explaining not only the emissions observed in the NGC\,7793~P13 spectrum, but also the relative velocities of the absorption components irrespective of the effective temperature or velocity law. Besides, our result is in good agreement with the value $\dot{M}\simeq10^{-5}\,M_{\odot}\,\text{yr}^{-1}$ by \cite{El_Mellah2019}, who estimated the minimum acceptable mass outflow rate in the Roche-lobe filling donor star wind that can still maintain the observed X-ray luminosity of the pulsar. 

The mass loss rate higher by over an order of magnitude than typical B supergiants have may be related to rapid rotation of the star, with a rotation velocity close to its critical value \citep{Maeder2000}. For instance, using formula (4.30) from \citet{Maeder2000} with the index $\alpha=0.20$\footnote{The quantity $\alpha=g^{\rm thick}_{\rm rad}/g^{\rm tot}_{\rm rad}$ characterizes the contribution of the optically thick lines $g^{\rm thick}_{\rm rad}$ to the total radiation pressure $g^{\rm tot}_{\rm rad}$.} for $\log{T_{\rm ph}}\approx4.00$ \citep{Lamers1995} and $\Gamma_{\rm es}\approx0.3$\footnote{$\Gamma_{\rm es}$ is the ratio of the acceleration of the wind matter induced by scattering of the central source radiation on free electrons to the free fall acceleration at the hydrostatic radius $R_*$.} from our model, one obtains a wind amplification factor of 5 for $v_{\rm rot}=0.5v_{\rm crit}$. At $v_{\rm rot}\approx v_{\rm crit}$, the possible mass loss rate increases up to 1.5--2 orders of magnitude compared to that of a non-rotating star.

This scenario ($v_{\rm rot}\approx v_{\rm crit}$) is believed to take place in B[e] supergiants (sgB[e]) whose winds exhibit two main components: a fast, hot polar wind and a cold, slow, more dense equatorial one \citep{Zickgraf1985} forming a disk-like structure. The wind density in the polar and equatorial planes of the extended envelope of B[e] supergiants differs by about one order of magnitude, so the mass loss rate in the star's disk may be even several times higher than $10^{-5}\,M_{\odot}\,\text{yr}^{-1}$ \citep{Zickgraf1992, Kraus2007}. The presence of a donor of this type in NGC\,7793~P13 could provide both the high mass loss rate in the polar wind necessary for describing the observed spectrum and more effective feeding of the neutron star compared to the case considered by \citet{El_Mellah2019}. 

In order to reach the high mass inflow rates into the accretion disk, it is preferable (but still not strictly necessary) for the neutron star to move inside a dense equatorial disk-like wind of the B[e] supergiant whose half-opening angle is about $10^\circ$ \citep{Zickgraf1992}, i.\,e. it is preferable for the equatorial plane of the star to coincide with the orbital plane within the mentioned angle. In this case, the accretion disk will also be in the same plane, but, as we noted in the previous section, donor heating assumes its falling into the X-ray radiation cone, the axis of which should therefore be tilted to the orbital plane by a rather significant angle (at least $10^\circ$--$30^\circ$ as was shown above). This implies that the funnel together with the inner regions of the accretion disk should be inclined with respect to its outer parts and the orbital plane.

Such a violation of the coplanarity of different parts of the disk may be related to, e.\,g., the effect of propagation of bending waves in viscous disks associated with the relativistic Lense–Thirring precession \citep{Bardeen1975,Ingram2009}. Although the resulting distortion of the disk shape depends on multiple parameters, the tilt angle difference between the inner and outer parts of the disk can potentially reach the required tens of degrees if the compact object's spin axis is highly tilted with respect to the orbital plane; this could happen as a result of the supernova explosion \citep{Fragos2010}. Actually, the Lense–Thirring precession has already been proposed by \citet{Middleton2018} as a possible explanation for a superorbital period of NGC\,7793~P13. Despite the fact that the authors considered the $P\approx65$~day period as superorbital, which is, apparently, orbital after all \citep{Furst2021}, the relations obtained by \citet{Middleton2018} can also explain longer periods. Another mechanism capable of causing precession in a deformed disk is the interaction of the disk matter with the dipole component of the neutron star magnetic field \citep{Pfeiffer2004}. In the case of strong $B\sim10^{14}$~G fields the precession period may reach years \citep{Mushtukov2017}, which is comparable with the $P\gtrsim1500$~day period known for NGC\,7793~P13 \citep{Motch2014,Hu2017,Furst2018}.

Similar to close binaris with Be stars \citep{Reig2011}, the disk of the B[e] supergiant may be severely trimmed by tidal forces. Considering the fact that the semi-major axis of the neutron star orbit in NGC\,7793~P13 is less than two donor radii, only a small part of the disk should survive. One of expected consequences of this is a significant decrease in intensities of the forming in the disk permitted and forbidden lines of metals that usually observed in sgB[e] stars \citep[see the review by][and references therein]{Kraus2019}. The rather small infrared excess in the NGC\,7793~P13 spectrum \citep{Lau2019} may also have the same reason. As a result, the spectrum of our hypothetical B[e] supergiant should become similar to the spectra of typical B supergiants, which is observed in the considered case.

The hypothesis of the presence of an sgB[e] star in the NGC\,7793~P13 system has two more consequences which, at first glance, may contradict the observations. For the obtained photospheric radius of the donor $105\,R_\odot$ and the star mass $18-23\,M_\odot$ \citep{Motch2014}, the upper estimate for the critical rotation velocity\footnote{The critical rotation velocity was calculated according to the relation (1) in \citet{Townsend2004}.} is about $v_{\rm crit}\approx180-200$~km\,s$^{-1}$. At the same time, the projected rotation velocity for the P13 donor measured from the absorption line widths, $v\sin{i}$ is less than 100~km\,s$^{-1}$ \citep{Motch2014}. This discrepancy can be resolved if one supposes a small inclination of both the star rotation axis and the orbital plane to the line of sight. Indeed, relatively small orbital inclinations of \mbox{$i\lesssim20$--$30^\circ$} do not contradict the results of \citet{Motch2014} and \citet{Furst2021}. Additionally, one should take into account the influence of gravitational darkening on the absorption line widths in rapidly rotating stars. This effect reduces the contribution of emissions forming in equatorial regions and, consequently, leads to a slower increase in the line widths with increasing rotation velocity in comparison with the models with a single photospheric temperature. Thus, the projected rotation velocity measured from the widths of absorption lines might be significantly underestimated, and the difference between the real and measured $v\sin{i}$ should increase as the photospheric temperature decreases, reaching 30\%\ or even more for late-type B stars \citep{Townsend2004}.

The second discrepancy is the fact that the obtained $v_{\rm crit}$ exceeds by over 30\%\ the rotation velocity $v_{\rm syn}$ that the donor should have in the case of its synchronized rotation with the neutron star in the periastron of the elliptical orbit\footnote{We compute it assuming the eccentricity $e=0.24$, \citet{Furst2021}.}; but the usually observed velocities are closer to $v_{\rm syn}$ \citep[see, e.g.,][]{Koenigsberger2012}. If the binary components in NGC\,7793~P13 have a circular orbit, this difference is even higher. On the other hand, \citet{Koenigsberger2012} reported $v_{\rm rot}/v_{\rm syn}= 1.2\pm0.1$ for Cyg~X-2, so NGC\,7793~P13 is not an unique case.

Finally, let us note that despite mechanisms of B[e] supergiant formation in binaries are not clear and remain a subject of discussion \citep{Langer1998, Kraus2019}, currently, there are only several cases of reliable detection of B[e] supergiants in pairs with stars of various types \citep[e.g.,][]{Miroshnichenko2002, Marchiano2012, Wheelwright2012, Clark2013}, and a few cases of not so confident detection in pairs with X-ray sources (including ULXs) still requiring confirmations, e.g., CI\,Cam \citep{Bartlett2019} or Holmberg\,II~X-1 \citep{Lau2017}.

\section{Conclusions}
In this work we have shown that most of the observed features of the optical spectrum of the ultraluminous X-ray pulsar NGC\,7793~P13 can be described within a spherically-symmetric donor star model with a wind mass loss rate of $\dot{M}=5.8\times10^{-6}\,M_{\odot}\,\text{yr}^{-1}$ after accounting for the effects of irradiation by X-ray quanta of both the star surface and the wind matter. This model allowed us to obtain the P\,Cyg hydrogen and neutral helium line profiles in the most natural way, as well as to reproduce the depth and position of the absorption lines, what we were unable to do with the two-component model considered earlier where we combined the spectra of a standard B9\,Ia star and an additional source of bright emissions, whose role was played by the outflow from the surface of the supercritical accretion disk. Some of the results and their consequences required a more detailed consideration: 
\begin{list}{}{
\setlength\leftmargin{5mm} \setlength\topsep{0mm}
\setlength\parsep{0mm} \setlength\itemsep{0mm} }
\item[1)] the luminosity of the X-ray irradiating source used for reproducing the He\,II~$\lambda4686$ emission intensity turned out to be lower by 2.5-3 orders of magnitude than the X-ray luminosity of the system in the assumption of isotropic emission; 
\item[2)] despite irradiation by the X-ray source with a luminosity of $10^{39}-10^{40}~$erg\,s$^{-1}$, the stellar wind is not suppressed, as would be expected based on the previous studies of X-ray binaries;
\item[3)] the mass loss rate in the donor wind is more than 10 times greater than values typical of standard B9\,Ia stars. 
\end{list}
We have shown that the first two problems can be qualitatively solved if we consider a moderate collimation of the X-ray emission instead of an isotropic source. This allows one to decrease the amount of actually emitted energy and limit the direction of X-ray quanta propagation in space, which is precisely what is required for survival of the donor star wind. The increased mass loss rate in the wind of the star may be related to its fast rotation, the velocity of which may be close to the critical level. We should note that all the suggestions put forward require numerical verification far beyond the scope of this paper.

\section*{Acknowledgments}
This work is based on observations collected at the European Organisation for Astronomical Research in the Southern Hemisphere under program 084.D-0881(A).

\section*{Funding}
This work was carried out with the financial support of the Russian science foundation (project No. 21-72-10167 «Ultraluminous X-ray sources: wind and donors»).

\section*{Conflict of Interest}
The authors declare no conflict of interest.

\end{document}

%% file: sao_cmd_author.tex
\def\squareforqed{\hbox{\rlap{$\sqcap$}$\sqcup$}}

\def\sq{\ifmmode\squareforqed\else{\unskip\nobreak\hfil
\penalty50\hskip1em\null\nobreak\hfil\squareforqed
\parfillskip=0pt\finalhyphendemerits=0\endgraf}\fi}

\def\utw{\smash{\rlap{\lower5pt\hbox{$\sim$}}}}

\def\udtw{\smash{\rlap{\lower6pt\hbox{$\approx$}}}}

\def\diameter{{\ifmmode\mathchoice
{\ooalign{\hfil\hbox{$\displaystyle/$}\hfil\crcr
{\hbox{$\displaystyle\mathchar"20D$}}}}
{\ooalign{\hfil\hbox{$\textstyle/$}\hfil\crcr
{\hbox{$\textstyle\mathchar"20D$}}}}
{\ooalign{\hfil\hbox{$\scriptstyle/$}\hfil\crcr
{\hbox{$\scriptstyle\mathchar"20D$}}}}
{\ooalign{\hfil\hbox{$\scriptscriptstyle/$}\hfil\crcr
{\hbox{$\scriptscriptstyle\mathchar"20D$}}}}
\else{\ooalign{\hfil/\hfil\crcr\mathhexbox20D}}%
\fi}}

\newcommand{\aap}{Astron. and Astrophys. }

\newcommand{\aj}{Astron.~J. }
\renewcommand{\apj}{Astrophys.~J. }
\newcommand{\apjs}{Astrophys.~J. Suppl. }
\newcommand{\apss}{Astrophys. and Space Sci. }

\newcommand{\mnras}{Monthly Notices Royal Astron. Soc. }

\newcommand{\pasj}{Publ. Astron. Soc. Japan }
\newcommand{\pasp}{Publ. Astron. Soc. Pacific }

\renewcommand{\nat}{Nature }

\newcommand{\apjl}{Astrophys.~J.}